\documentclass[11pt,twoside]{article}

\usepackage{asp2006}
\usepackage{epsf}
\usepackage{lscape}
\usepackage{graphicx}
\usepackage[]{natbib}
\usepackage[usenames]{color}

\bibpunct{(}{)}{;}{a}{}{,} 
\def\kms    {\ifmmode{{\rm ~km\,s}^{-1}}\else{~km\,s$^{-1}$}\fi}

\def\hi {HI}
\def\mhi {M$_{HI}$}

\begin{document}
\title{Setting the normalcy level of HI properties in isolated galaxies}   
\author{D. Espada$^{1,2}$,
           L. Verdes-Montenegro$^{1}$,
	    E. Athanassoula$^{3}$,
           A. Bosma$^{3}$,
           W. K. Huchtmeier$^{4}$,
           S. Leon$^{5}$, 
           U. Lisenfeld$^{6}$, 
           J. Sabater$^{1}$,
           J. Sulentic$^{1}$,
           S. Verley$^{6}$ and
           M. Yun$^{7}$ }   
\affil{$^{1}$ Instituto de Astrof\'{i}sica de Andaluc\'{i}a (IAA-CSIC) (daniel@iaa.es); $^{2}$
 Harvard-Smithsonian Center for Astrophysics (CfA); $^{3}$ Laboratoire d'Astrophysique de Marseille; $^{4}$ Max-Planck-Institut fuer Radioastronomie (MPIfR); $^{5}$ Instituto de Radioastronom\'{i}a Milim\'{e}trica (IRAM); $^{6}$ Universidad de Granada; $^{7}$ UMass-Amherst}    

\begin{abstract} 
Studying the atomic gas (HI) properties of the most isolated galaxies is essential  to quantify the effect that the environment exerts on this sensitive  component of the interstellar medium.  We observed and compiled HI data for a  well defined sample of $\sim$ 800 galaxies in the Catalog of Isolated Galaxies, as part of the AMIGA project (Analysis of the ISM in Isolated GAlaxies, \verb!http://amiga.iaa.es!), which enlarges considerably previous samples used to quantify the HI deficiency in galaxies located in denser environments. By studying the shape of 182 HI profiles, we revisited the usually accepted result that, independently of the environment, more than half of the galaxies present a perturbed HI disk. In isolated galaxies this would certainly be a striking result if these are supposed to be the most relaxed systems, and has implications in the relaxation time scales of HI disks and the nature of the most frequent perturbing mechanisms in galaxies.  Our sample likely exhibits the lowest HI asymmetry level in the local Universe. We found that field samples present an excess of $\sim$ 20\% more asymmetric HI profiles than that in CIG. Still a small percentage of galaxies in our sample present large asymmetries. Follow-up high resolution VLA maps give insight into the origin of such asymmetries.

\end{abstract}

\vspace{-0.8cm}

\section{Introduction}

Atomic gas (\hi ) is presumably the most extended cold
component  of the interstellar medium (ISM) and is very sensitive to any kind of perturbation. 
For instance, the \hi\ component in spiral galaxies has long been known to show both  geometric and kinematic
asymmetries \citep[e.g.][]{1969Natur.221..531B,1980MNRAS.193..313B}, 
and the HI content is found to be deficient in galaxies in denser environments \citep[e.g.][]{1973MNRAS.165..231D}.
Although asymmetries have also been seen in stellar disks \citep[e.g.][]{1997ApJ...477..118Z}, the fact that HI is usually twice more extended than the optical component makes it arguably a better tracer of interaction. 
The study of \hi\ global velocity profiles of galaxies has proven to be very useful for both 
the quantification of the HI content as well as the asymmetry level.  In the context of the AMIGA project, we use \hi\ global velocity profiles of a large sample of strictly isolated galaxies in order to: $i)$ revise the reference to study the HI content normalcy level  in galaxies, $ii)$ quantify the rate of \hi\ asymmetries, and $iii)$ study the relevance of different mechanisms producing asymmetries in galaxies, with the additional help of aperture synthesis maps.

\vspace{-0.3cm}
\section{HI content}

We present \hi\ data  for $N$ = 834 CIG galaxies (Catalog of Isolated Galaxies, \citealt{1973AISAO...8....3K}) and study the correlations between the \hi\ content and different parameters associated with the stellar component: optical luminosities (see Figure~1, left), diameters and morphological type. We calculated \hi\ fluxes, mean velocities, line-widths and \hi\ masses for galaxies we observed at the Arecibo, Effelsberg, Nan\c{c}ay and Green Bank (GBT) radio-telescopes, or compiled  from different sources in the literature  \citep{2009E}. Survival analysis was applied in order to use the information given by upper limits.
  This sample is about 3 times larger than that of \citet{1984AJ.....89..758H}, the prevalent reference of HI data for isolated galaxies to date, increasing considerably the population per morphological bin. This is important especially for the less populated bins, such as E and Sd/Irr, where our sample has a factor 5--10 more galaxies.
Our improved sample is thus essential as a control sample for studies involving galaxies in denser environments.
We note that to compare the HI content of two different samples, one needs to homogeneously adopt the same constants and corrections, since otherwise the differences may be comparable to the observed levels of deficiency. 

\vspace{-0.3cm}
\section{Global measure of HI asymmetry}

 The frequency and amplitude of gaseous disk asymmetries have been studied by several authors, with the main conclusion that at least half of the galaxies are asymmetric even in the field \citep[e.g.][]{1994A&A...290L...9R}. 
We restricted our study to a subsample of $N$ = 182 galaxies for which we have  global \hi\ velocity profiles with $S/N$ $>$ 10 and minimized undesired artificial lopsidedness. We calculated a flux ratio asymmetry parameter to quantify the asymmetry of the profiles, defined as $A_{flux~ratio}= A_{l/h}$, if $A_{l/h}$ $>$ 1, and
$1/A_{l/h}$  otherwise,  where $A_{l/h}$ is the ratio of the areas under the profile at velocities lower and higher than
the central velocity (e.g. \citealt{1998AJ....115...62H}). We minimized artificial asymmetries by avoiding large pointing offsets and poor determinations of the central velocity which resulted in possible errors of $A_{flux~ratio}$ larger than 10\%.

\begin{figure}
\begin{center}
\includegraphics[width=8cm]{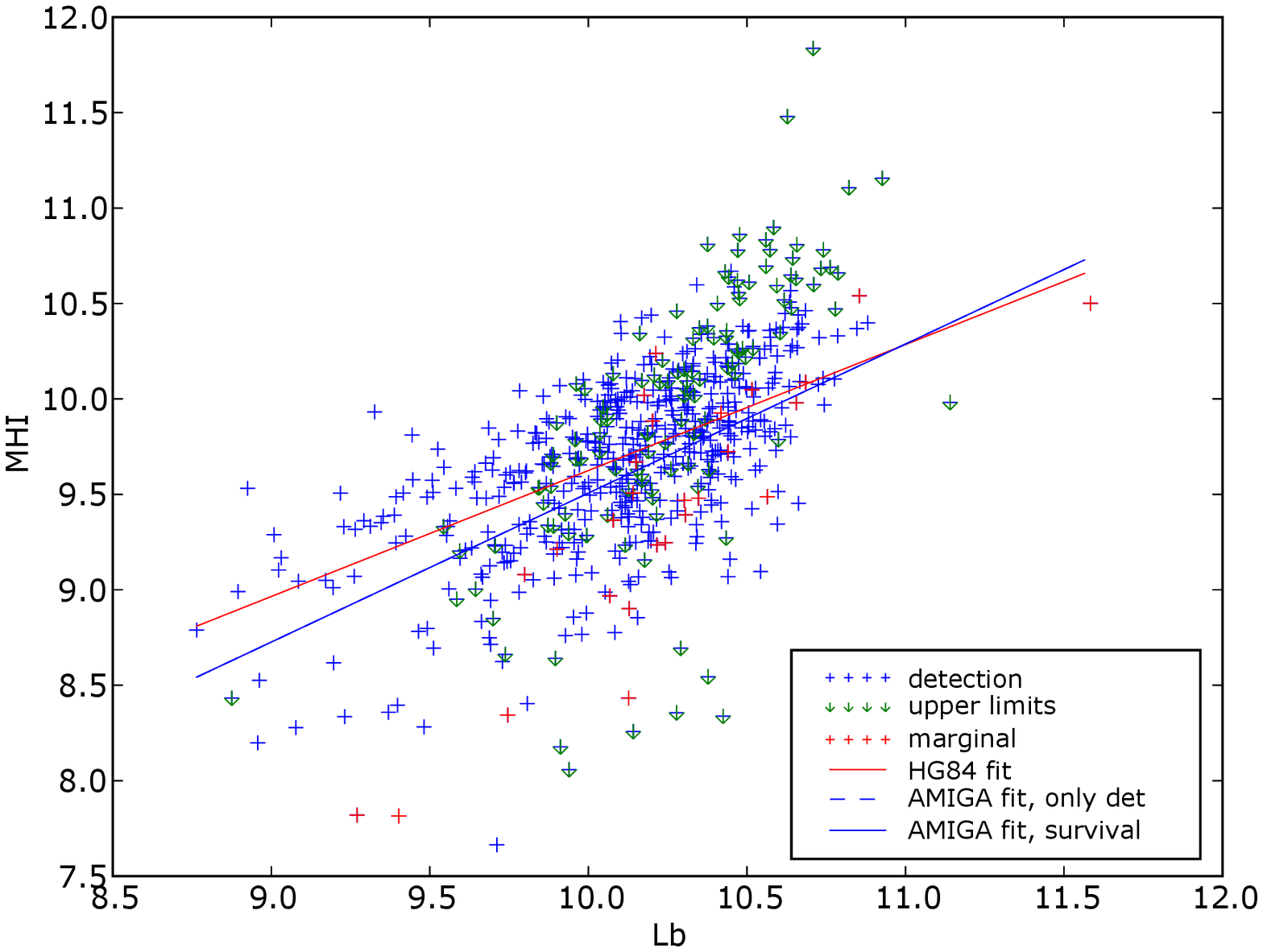}
\includegraphics[width=5cm]{despada_fig1b.eps}
\caption{$(Left):$ \mhi\ vs $L_B$ plot for the whole sample of $N$ = 837 galaxies. Detections (cross signs) as well as upper limits (arrow signs) are included. The fit presented by Haynes \& Giovanelli (1984) is shown as a red line, and that of our sample in blue.
$(Right):$ $A_{flux~ratio}$ distribution, half-Gaussian fit and the residuals. 
\label{fig1}}
\end{center}
\end{figure}

The resulting $A_{flux~ratio}$ distribution was characterized by a half-Gaussian profile with a FWHM/2 of 0.13 (Figure~\ref{fig1}, right). 
This width is very likely smaller than 0.13, since remaining systematic errors were estimated to be $\sim$ 7\%, mostly produced by pointing offsets and an inaccurate determination of the mean velocity.
For large values of the parameters, $A_{flux~ratio}$ $>$ 1.3, we found a total of 16 galaxies. This corresponds to an excess of 12 more than what would correspond to the fitted half-Gaussian profile, or a  6\% of the subsample. This is very likely a residual of interacting systems.

\begin{figure}
\begin{center}
\hspace{-2cm}
\includegraphics[height=5.3cm,width=6.5cm]{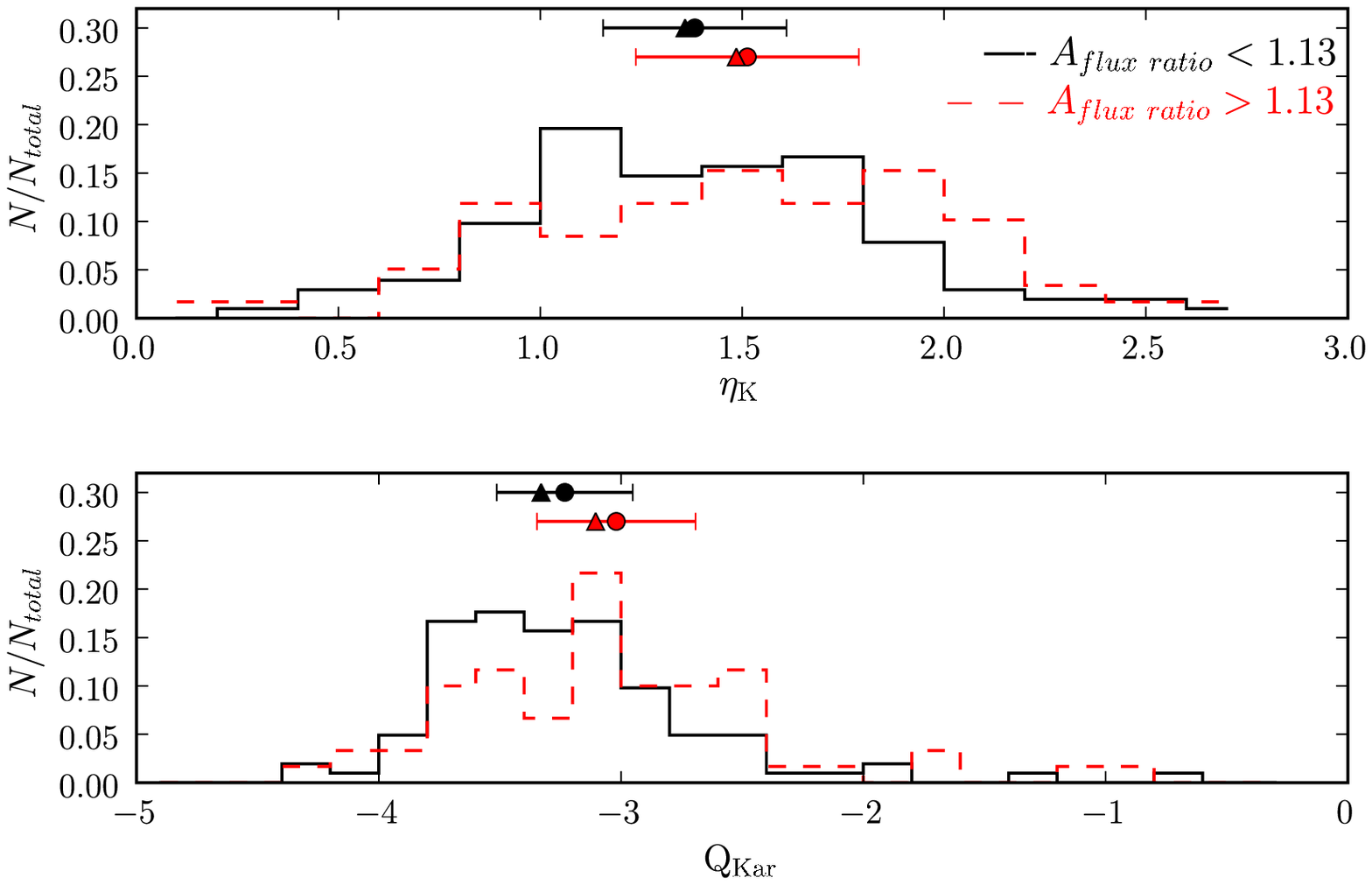}
\includegraphics[width=8.5cm]{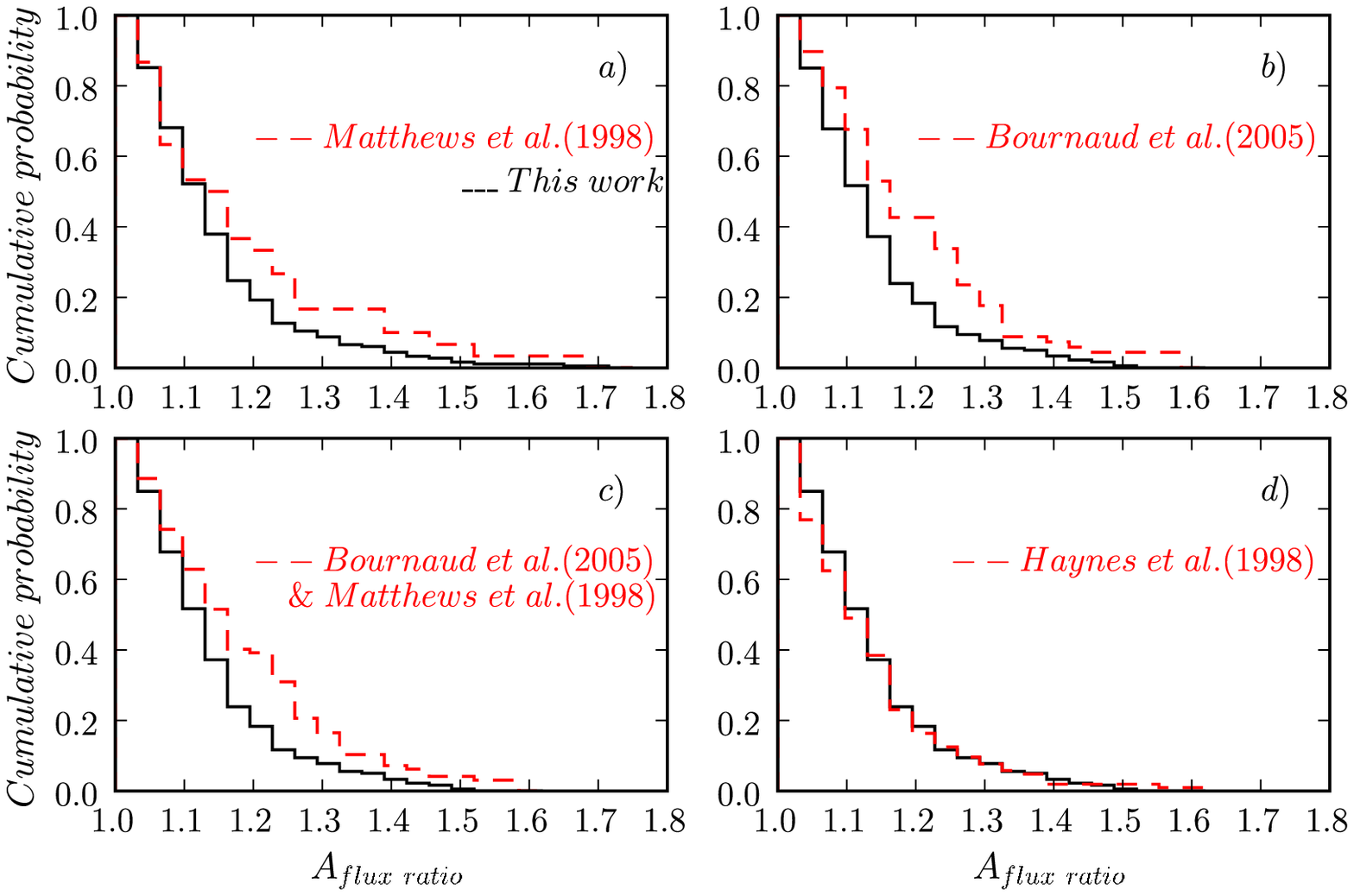}
\hspace{-2cm}
\caption{$(Left):$ Isolation parameter distribution ($Q_{Kar}$ and $\eta_K$) for symmetric ($A_{flux~ratio}$ $<$ 1.13, solid-line histogram) and asymmetric ($A_{flux~ratio}$ $>$ 1.13, dashed-line histogram) HI profiles.  
   The mean, standard deviation and median are shown as  circle, bar and triangle signs, respectively.
 $(Right):$  Comparison of  the cumulative  $A_{flux~ratio}$ distribution between our sample (solid line) and other samples (dashed line). 
 }
\label{fig2}
\end{center}
\end{figure}

Among our sample of  isolated galaxies we find a weak trend for larger \hi\ asymmetries to be related to less isolated systems.  We plot in Figure~\ref{fig2} (left) the comparison of the distribution of
 the tidal strength ($Q_{Kar}$) and galaxy surface number density ($\eta_K$) parameters (\citealt{2007A&A...470..505V}$a$,$b$) for the most symmetric  ($A_{flux~ratio}$ $<$ 1.13) and the most asymmetric profiles ($A_{flux~ratio}$ $>$ 1.13). Smaller $\eta_K$ and $Q_{Kar}$ are found for galaxies with symmetric \hi\ profiles. 
  


In  Figure~\ref{fig2} (right) we show the $A_{flux~ratio}$ cumulative probability distribution of previous samples of field/isolated galaxies and compare it with that in our sample. We found a $\sim$ 20\% excess of  asymmetric profiles in samples of field galaxies or not obviously interacting galaxies. The only similar distribution to ours is that of Haynes et al. (1998), whose sample comprises some isolated galaxies, although not members of CIG. 
The rate of asymmetric profiles is expected to be even higher in strongly interacting galaxies.

\vspace{-0.4cm}

\section{Origin of HI asymmetries}

We carried out a follow-up study of the origin of the \hi\ asymmetries for eight galaxies with the most asymmetric profiles as well as four galaxies with the most symmetric \hi\ profiles for comparison, using VLA aperture synthesis \hi\ observations.
One of the isolated galaxies with an asymmetric profile in this subsample is CIG~96 (NGC~864), whose \hi\ synthesis imaging from the VLA has been studied in detail in \citet{2005A&A...442..455E}. The asymmetry in its \hi\ profile is associated with a strong kinematical perturbation in the gaseous envelope of the galaxy, where on one side the decay of the rotation curve is faster than Keplerian. Although a companion is detected, no tidal tail is found, and it is probably not massive enough to cause such a perturbation. Probably we are witnessing the capture of a gaseous companion.
Aperture synthesis maps for the rest of the sources showed more symmetric structures, both in geometry and kinematics. Neither tidal tails nor relevant HI rich companions were found close to any of the galaxies. The asymmetry was usually found in the kinematics, in the sense that the rotation curve on one side of the galaxy was steeper.
These results are similar to that in other galaxies \citep{1999MNRAS.304..330S}, not necessarily isolated. 
  Given that in average galaxies in our sample did not likely undergo a major
   interaction since less than 3~Gyr ago \citep{2005A&A...436..443V},
   these asymmetries may either be the remnants of a major
   interaction longer ago, or of a more recent short-lived mechanism such as a minor merger.

{\bf Acknowledgments.} This work has been partially supported by Spanish DGI grant  AYA2008-06181-C02, as well as Junta de Andalucia (Spain) grants P08-FQM-4205-PEX and TIC-114 16-C02-01.
DE is supported by the Marie Curie International fellowship MOIF-CT-2006-40298.

\end{document}